%% file: sn-article.tex
\documentclass[pdflatex,sn-mathphys-num]{sn-jnl}

\usepackage{amsmath,amssymb,amsfonts}
\usepackage{algorithmic}
\usepackage{graphicx}
\usepackage{textcomp}
\usepackage{listings}
\usepackage{float}
\usepackage{multirow}
\usepackage{hyperref}
\usepackage{soul}
\usepackage[ruled,vlined, linesnumbered]{algorithm2e}
\usepackage{caption}
\usepackage{subcaption}
\usepackage{tcolorbox}
\usepackage{xcolor}   

\lstdefinestyle{mystyle}{
    language=Python,
    backgroundcolor=\color{white},   
    commentstyle=\color{codegreen},
    keywordstyle=\color{magenta},
    numberstyle=\tiny\color{codegray},
    stringstyle=\color{codepurple},
    basicstyle=\ttfamily\footnotesize,
    breakatwhitespace=false,         
    breaklines=true,                 
    captionpos=b,                    
    keepspaces=true,                 
    numbers=left,                    
    numbersep=5pt,                  
    showspaces=false,                
    showstringspaces=false,
    showtabs=false,                  
    tabsize=2
}
\definecolor{codegreen}{rgb}{0,0.6,0}
\definecolor{codegray}{rgb}{0.5,0.5,0.5}
\definecolor{codepurple}{rgb}{0.58,0,0.82}

\theoremstyle{thmstyleone}%
\theoremstyle{thmstyletwo}%

\theoremstyle{thmstylethree}%

\usepackage{etoolbox}
\makeatletter
\patchcmd{\@maketitle}{Contributing authors:}{}{}{}
\makeatother

\author[1]{\fnm{Yuxuan} \sur{Wang}}
\author[1]{\fnm{Jingshu} \sur{Chen}}
\author[2]{\fnm{Qingyang} \sur{Wang}}

\affil[1]{\orgdiv{Department of Computer Science}, \orgname{University of Alabama in Huntsville}, \city{Huntsville}, \state{AL}, \country{USA}}

\affil[2]{\orgdiv{Department of Computer Science and Engineering}, \orgname{Louisiana State University}, \city{Baton Rouge}, \state{LA}, \country{USA}}

\email{\textsuperscript{1}\{yw0029, jc1540\}@uah.edu}
\email{\textsuperscript{2}qwang26@lsu.edu}

\begin{document}

\title{Leveraging Large Language Models for Command Injection Vulnerability Analysis in Python: An Empirical Study on Popular Open-Source Projects}








\abstract{Command injection vulnerabilities are a significant security threat in dynamic languages like Python, particularly in widely used open-source projects where security issues can have extensive impact. With the proven effectiveness of Large Language Models(LLMs) in code-related tasks, such as testing, researchers have explored their potential for vulnerabilities analysis. This study evaluates the potential of large language models (LLMs), such as \texttt{GPT-4}, as an alternative approach for automated testing for vulnerability detection. In particular, LLMs have demonstrated advanced contextual understanding and adaptability, making them promising candidates for identifying nuanced security vulnerabilities within code. To evaluate this potential, we applied LLM-based analysis to six high-profile GitHub projects—Django, Flask, TensorFlow, Scikit-learn, PyTorch, and Langchain—each with over 50,000 stars and extensive adoption across software development and academic research. Our analysis assesses both the strengths and limitations of LLMs in detecting command injection vulnerabilities, evaluating factors such as detection accuracy, efficiency, and practical integration into development workflows. In addition, we provide a comparative analysis of different LLM tools to identify those most suitable for security applications. Our findings offer guidance for developers and security researchers on leveraging LLMs as innovative and automated approaches to enhance software security.}




\maketitle

\input{introduction}
\input{background}
\input{dataset}
\input{approach}
\input{evaluation}
\input{related_work}

\input{threats_to_validity}
\input{conclusion}

\onecolumn
\appendix
\input{appendix}

\end{document}

%% file: introduction.tex
\section{Introduction}

Python's rich ecosystem makes it an ideal choice for a wide range of AI tasks, from scripting to developing complex models. Widely adopted libraries, such as TensorFlow \citep{tensor} and PyTorch \citep{pytorch}, offer robust tools for machine learning and deep learning applications. However, as the use of AI techniques and related open-source software continues to grow, addressing security vulnerabilities becomes increasingly essential—particularly in popular libraries and frameworks where security flaws can lead to widespread, systemic risks. 

One such security vulnerability, command injection, represents a critical threat in dynamic languages like Python. Command injection vulnerabilities allow attackers to exploit applications by executing unauthorized commands, potentially compromising system integrity and exposing sensitive data. An exploited vulnerability can lead to data breaches, unauthorized access to sensitive resources, and loss of system control. For instance, an attacker could access confidential data or manipulate system files, which can disrupt service and harm user trust \citep{stasinopoulos2019commix}. The urgency of addressing these vulnerabilities is underscored by their prominent ranking in security advisories, such as the Common Weakness Enumeration (CWE) \citep{CWE}, and by recent alerts \citep{civ_fbi} from organizations like CISA and the FBI, which highlight the risks these vulnerabilities pose to common software products.

The Common Vulnerabilities and Exposures (CVE) database has documented multiple instances of command injection vulnerabilities within Python libraries \citep{CVE}. One such example, shown in Listing \ref{instance}, involves a vulnerability reported in CVE-2022-29216 \citep{CVE22}. This vulnerability existed in the \texttt{preprocess\_input\_exprs\_arg\_string} function within the \texttt{saved\_model\_cli.py} file, where an \texttt{eval()} method call enabled command injection through unvalidated inputs. By passing malicious commands via the \texttt{input\_exprs\_str} parameter, an attacker could exploit this function due to the lack of an input validation mechanism.

Existing tools, such as Bandit \citep{bandit}, have been instrumental in identifying certain types of command injection vulnerability. However, those static analysis tools often require fully compiled code and may not adapt well to the nuances of large, fragmented codebases or evolving vulnerability patterns. Recent advances in large language models (LLMs), including models like GPT-4, offer promising alternatives in code analysis and generation \citep{ouyang2023llm, dai2023llm, 10329992, tang2023chatgpt, 10.1145/3624032.3624035}. Unlike traditional tools, LLM-based approaches could analyze both compiled and non-compiled code fragments and generate contextually relevant vulnerability assessments without requiring the complete code structure. In addition, an LLM-based approach can generate security tests that validate vulnerability assessments, providing an additional layer of security assurance.

For instance, as demonstrated in Section \ref{sec:moti}, detecting vulnerabilities in the code snippet in Listing \ref{realcase} presents motivation for our work in this paper. First, like many similar code examples, the snippet is fragmented and non-compilable, which limits the applicability of traditional vulnerability detection tools that rely on fully compiled code for analysis. Unlike conventional tools, large language models (LLMs) can analyze vulnerabilities directly within code snippets, regardless of their compilability. Furthermore, as demonstrated by the different analysis results of the two methods in section \ref{sec:moti1} and Section \ref{sec:moti2}, the approach of adopting \texttt{GPT-4} are capable of identifying vulnerabilities that may be overlooked by existing detection tools like \texttt{Bandit}. Additionally, LLMs can generate security tests for functions to validate their assessments, adding a layer of verification. These advantages motivate our proposed work of applying LLM-based approach to detect command injection vulnerabilities.

\lstset{xleftmargin=0.5cm}
\begin{lstlisting}[caption={Command injection vulnerability instance in Tensorflow.},label=instance]
 def preprocess_input_exprs_arg_string(input_exprs_str, safe=True):
  input_dict = {}

  for input_raw in filter(bool, input_exprs_str.split(';')):
    if '=' not in input_exprs_str:
      raise RuntimeError('--input_exprs "%s" format is incorrect. Please follow'
                         '"<input_key>=<python expression>"' % input_exprs_str)
    input_key, expr = input_raw.split('=', 1)
    if safe:
      try:
        input_dict[input_key] = ast.literal_eval(expr)
      except Exception as exc:
        raise RuntimeError(
            f'Expression "{expr}" is not a valid python literal.') from exc
    else:     
      input_dict[input_key] = eval(expr)  
  return input_dict

\end{lstlisting}

\textbf{Contributions:} 
The contributions of this paper are as follows:
\begin{itemize}
\item We conducted a comprehensive analysis of command injection vulnerabilities across 13,037 Python files from 6 most popular open-source projects, each with over 50K stars on GitHub. Our work explores the effectiveness and completeness of large language model (LLM)-based analysis for detecting command injection vulnerabilities, providing a rigorous evaluation of four different LLMs, including GPT-4 \citep{GPT4}, GPT-4o \citep{GPT4o}, Claude 3.5 Sonnet \citep{Claude3.5Sonnet} and DeepSeek-R1\citep{deepseek}.
\item We examined the characteristics of the command injection vulnerabilities that might be missed by LLM-based approach. Our results show the limitations of LLMs in this area and inspire future research in the domain
\item We compared our LLM-based approach to traditional security tools, specifically Bandit, to evaluate improvements in accuracy and completeness when identifying command injection vulnerabilities.
\item Our study built a dataset of 13,037 Python files from six high-profile GitHub projects—Django, Flask, TensorFlow, Scikit-learn, PyTorch, and Langchain—each with over 50,000 stars and extensive adoption across software development and academic research. This dataset is available in the GitHub, forming a benchmark resource for further research in vulnerability detection in the domain.

\end{itemize}

%% file: background.tex
\section{Motivation and Background}
\subsection{Motivating Example} \label{sec:moti}

In Listing \ref{realcase}, we present a Python function from the PyTorch \citep{pytorch} project as a motivating example. This function’s purpose is to retrieve and return a list of process IDs for all child processes associated with a given process ID (pid). In the second line, it employs the \texttt{subprocess.Popen()} method to execute the command \texttt{pgrep -P \{pid\}}. Among the parameters passed to \texttt{subprocess.Popen()}, \texttt{pid} is a formal parameter whose value can be controlled externally, and \texttt{shell=True} enables the execution of command strings directly in the shell. This combination presents a significant vulnerability to command injection attacks.To investigate its vulnerability, we applied two different analysis methods to this function: the existing detection tool, Bandit, and an LLM-based approach using GPT (e.g., \texttt{GPT-4}), and then compared their detection results.

\begin{lstlisting}[language=Python, caption={A candidate function from Pytorch project\cite{pytorch}.},label=realcase]
def get_child_pids(pid):
    pgrep = subprocess.Popen(args=f"pgrep -P {pid}", shell=True, stdout=subprocess.PIPE)
    pgrep.wait()
    out = pgrep.stdout.read().decode("utf-8").rstrip().split("\n")
    pids = []
    for pid in out:
        if pid:
            pids.append(int(pid))
    return pids
\end{lstlisting}

\subsubsection{Applying \texttt{Bandit} to Analyze the Function in Listing \ref{realcase}}\label{sec:moti1}

\begin{figure}[ht]
    \centering
    \begin{subfigure}{0.45\textwidth}        
        \includegraphics[width=\textwidth]{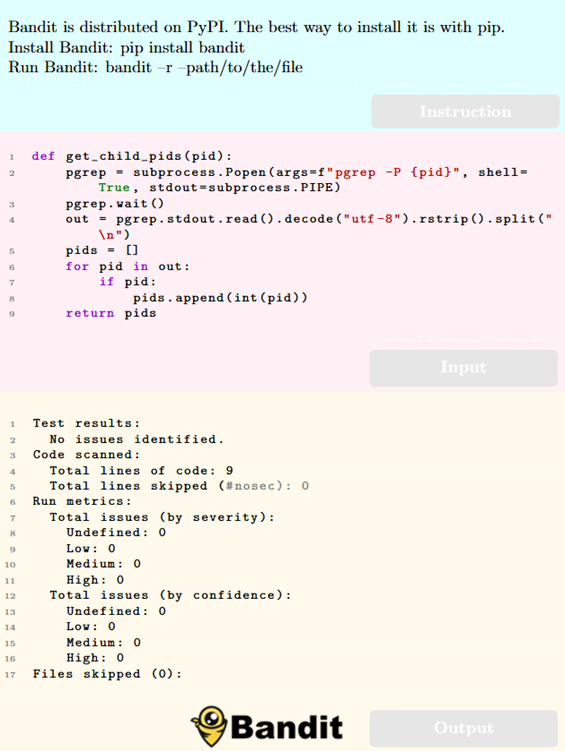}      
        \captionsetup{justification=centering}
        \caption{Result of Bandit}
        \label{banditprocess}
    \end{subfigure}
    \hspace{0.05\textwidth}
    \begin{subfigure}{0.45\textwidth}
        \includegraphics[width=\textwidth]{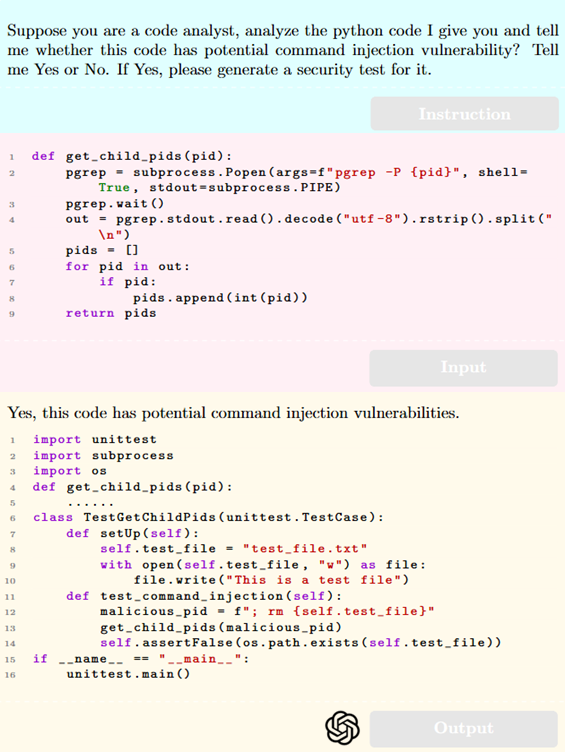} 
        \captionsetup{justification=centering}
        \caption{Result of GPT-4}
        \label{LLMprocess}
    \end{subfigure}
    \caption{The detection results for GPT-4 and Bandit.}
    \label{two detection results}
\end{figure}

First, we used Bandit \citep{bandit}, an existing tool for detecting security issues in Python code, to determine if there was a command injection vulnerability in this function. Figure \ref{banditprocess} illustrates the workflow for detecting code security issues with Bandit. We executed Bandit following the instructions shown in the blue block, while the yellow block displays Bandit's detection report. According to the report, Bandit concluded that this function does not contain a command injection vulnerability.

\subsubsection{Applying \texttt{GPT-4} to Analyze the Function in Listing \ref{realcase}} \label{sec:moti2}

Figure \ref{LLMprocess} illustrates our approach in using GPT-4 to analyze the function in Listing \ref{realcase}. The blue square contains the task assigned to GPT-4, while the yellow square displays GPT-4's output. The analysis begins with GPT-4 evaluating the function to identify any command injection vulnerabilities. As shown in Figure \ref{LLMprocess}, GPT-4's output indicates that it detected a command injection vulnerability in this function. Furthermore, GPT-4 generates a security test to verify the function's safety, adding confidence to its findings.

Figure \ref{LLMprocess} shows the security test generated by GPT for the get-\_child\_pids function. It is written using Python's unittest framework. In this code, a test file named test\_file.txt is first created for later testing. Next, a malicious command is passed to the get-\_child\_pids function, and since there is no protection, "pgrep -P \{pid\}" is interpreted as two commands: first execute pgrep -P, and then execute "rm \{self.test\_file\}", which deletes the test file.  After running this security test, the result obtained is True, which indicates that the test file has been deleted, means that the get\_child\_pids function is vulnerable to command injection attack. This result proves the analysis of the GPT-4 model.

\textbf{Motivation.} Detecting vulnerabilities in the code snippet shown in Listing \ref{realcase} motivates our work. The snippet, like many similar examples, is fragmented and non-compilable, which limits the effectiveness of traditional detection tools that require fully compiled code. In contrast, large language models (LLMs) can analyze vulnerabilities directly within code snippets. As demonstrated in Sections \ref{sec:moti1} and \ref{sec:moti2}, our approach using \texttt{GPT-4} identifies vulnerabilities that tools like \texttt{Bandit} may miss. Moreover, LLMs can generate security tests for functions to validate their assessments, adding an extra layer of verification. These advantages motivate our use of LLMs to detect command injection vulnerabilities.

\subsection{Large Language Models (LLMs)}

Large Language Models (LLMs) \citep{LLMs} are large-scale deep learning models designed to perform a wide range of natural language processing (NLP) tasks, including text recognition, translation, prediction, and generation. LLMs are primarily built upon the Transformer architecture \citep{AN2017NIPS}, employing self-supervised and semi-supervised learning methods to pre-train on extensive datasets, which allows them to learn language patterns and complex semantic structures. 

Due to their foundation on the Transformer, LLMs inherit its encoder-decoder structure and can generally be categorized into three main groups: encoder-only, encoder-decoder, and decoder-only models \citep{HZ2023arXiv}. Encoder-only models focus on understanding and encoding the input data. Examples include BERT \citep{BERT} and its variations, such as CodeBERT \citep{CodeBERT} and ALBERT \citep{ALBERT}. Encoder-decoder models utilize both encoding and decoding layers for tasks that require both input processing and output generation, with prominent examples being T5 \cite{T5} and CoTexT \citep{CoTexT}. Decoder-only models are primarily used for generation tasks; examples include the GPT model family, such as GPT-3 \citep{GPT3}, GPT-3.5 \citep{GPT3.5}, GPT-4 \citep{GPT4}, GPT-4o \citep{GPT4o} as well as other models like Google's PaLM \citep{PaLM}, Meta's LLaMA \citep{LLaMA} and DeepSeek-R1\citep{deepseek}. For Claude 3.5 Sonnet \citep{Claude3.5Sonnet}, a recently popular large language model, it cannot be determined at this time which architecture category it belongs to, as Anthropic has not released information about its internal architecture.

LLMs have demonstrated outstanding performance across various fields, including software engineering \citep{AA2022CCAI, CL2023arxiv, ZW2023arxiv, LZ2023arxiv, CT2023arxiv} and healthcare \citep{TA2023NM, VC2022arXiv}, where they have been widely adopted and rigorously evaluated. In this paper, we leverage GPT-4, GPT-4o, Claude 3.5 Sonnet and DeepSeek-R1, four state-of-the-art LLMs, for our experiments, given their advanced capabilities in understanding and generating complex code patterns, which are essential for effective vulnerability analysis.

%% file: dataset.tex
\section{Study Design}
\subsection{Research Questions} \label{sec:rq}

The aim of this study is to answer the following research questions:
\begin{itemize}
    \item \textbf{RQ1:} How effective are large language models (LLMs) like GPT in identifying command injection vulnerabilities in dynamic languages, specifically Python?
    \item \textbf{RQ2:} What types of Python command injection vulnerabilities might our LLM-based approach fail to detect?
    \item \textbf{RQ3:} What is the running cost(in terms of time and finance) of GPT-4?
    \item \textbf{RQ4:} How do different large language models compare in terms of accuracy, consistency, and efficiency in detecting command injection vulnerabilities in Python code?
    \item \textbf{RQ5:} How does the accuracy and efficiency of LLM-based vulnerability detection compare with traditional vulnerability analysis tools, such as Bandit?

\end{itemize}

\subsection{Dataset} \label{sec:data}

We selected six popular open-source Python projects, each with over 50,000 stars on GitHub, for our study. Table \ref{projects} lists the project names, versions, the total number of files, and the number of Python files in each project. These projects are widely used in academic research and software development, covering areas such as web framework development—Django \citep{django} and Flask \citep{flask}; machine learning modeling—TensorFlow \citep{tensor}, Scikit-learn \citep{scikit}, and PyTorch \citep{pytorch}; and LLM application development—Langchain \citep{langchain}. We filtered all Python files in these projects because our research focuses on detecting command injection vulnerabilities in Python code

\begin{table}[ht]
\caption{Studied projects(As Of January 2024)}
\label{projects}
\begin{tabular}{ccccccc}
\hline
Project      & Version  & \begin{tabular}[c]{@{}c@{}}No. of\\ Files\end{tabular} & Stars(K) & \begin{tabular}[c]{@{}c@{}}No. of \\ Python Files\end{tabular} & \begin{tabular}[c]{@{}c@{}}LOC of \\ Python Files(k)\end{tabular} & \begin{tabular}[c]{@{}c@{}}No. of \\ Candidate \\Functions\end{tabular} \\ \hline
Django       & 4.2.7    & 6,740                                                  & 80.5     & 2,772                                                          & 399.8                                                             & 17                                                                    \\ 
Flask        & 3.0.0    & 249                                                    & 68       & 82                                                             & 13.4                                                              & 2                                                                     \\ 
Langchain    & v0.0.330 & 3,933                                                  & 94.5     & 2,290                                                          & 227.1                                                             & 13                                                                    \\ 
TensorFlow   & 2.14.0   & 31,082                                                 & 186      & 3,106                                                          & 1,014.7                                                           & 46                                                                    \\ 
Scikit-learn & 1.3.2    & 1,569                                                  & 60       & 923                                                            & 317.5                                                             & 7                                                                     \\ 
PyTorch      & 2.1      & 12,401                                                 & 83.7     & 3,864                                                          & 1,234                                                             & 105                                                                   \\ \hline
\end{tabular}
\end{table}

%% file: approach.tex
\section{Our Approach}

\subsection{Approach Overview}

\begin{figure}[ht]
  \centering
  \includegraphics[width=\textwidth]{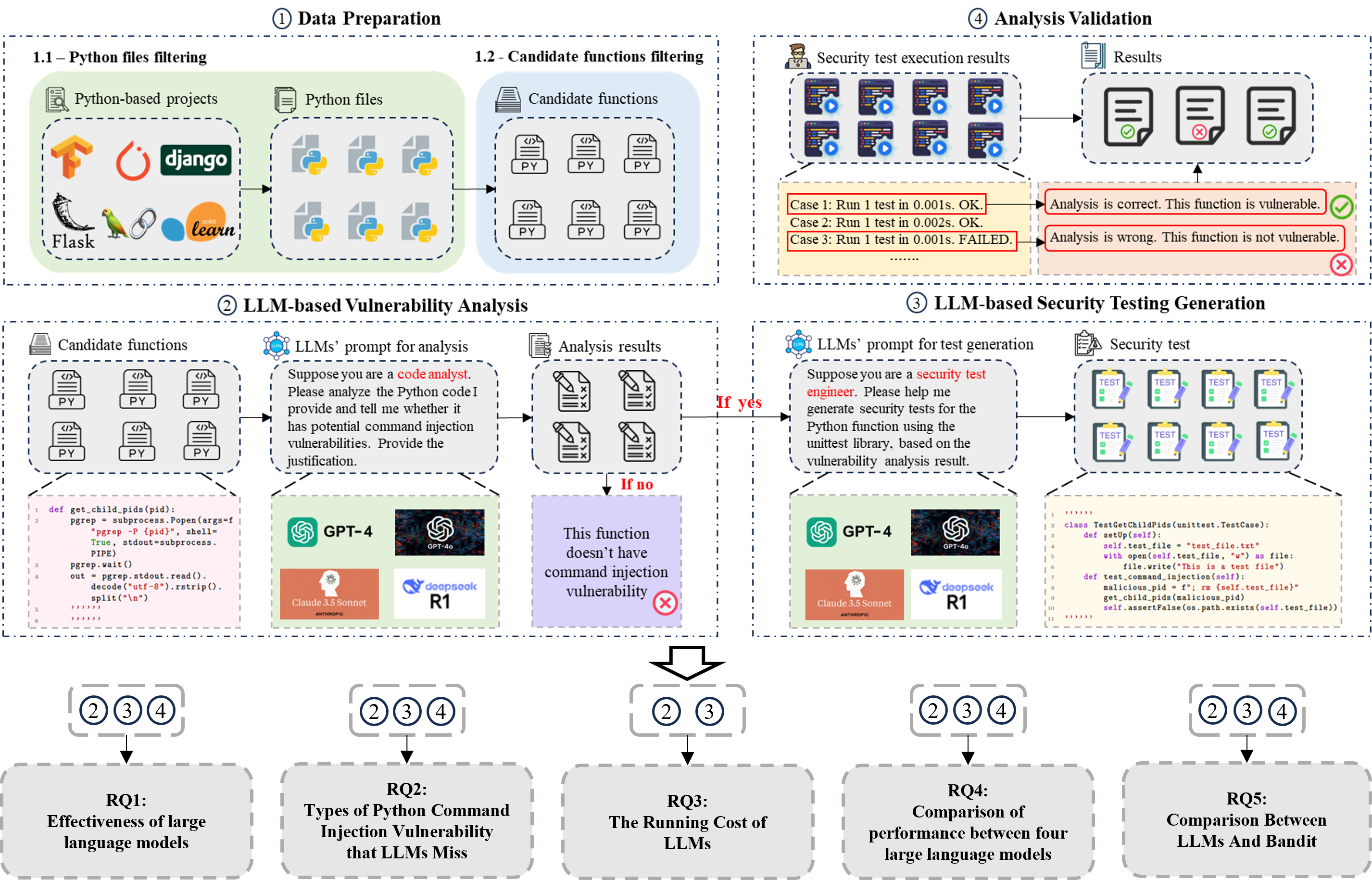}
  \caption{The overview of the proposed LLM-based approach.}
  \label{process}
\end{figure}

 Figure \ref{process} illustrates the overview of our approach workflow, which consists of five steps: (1) filtering Python files from the projects, (2) identifying candidate functions, (3) using LLMs to detect vulnerabilities, (4) using LLMs to generate security tests, and (5) validating the analysis results of LLMs. In the remaining section, we present each step in details.
 
\subsection{Data preparation}
\label{1ststep}

In the six target Python projects, in addition to the \texttt{.py} source files, there are image resources, README documents, and various other non-Python files. Since our objective is to detect command injection vulnerabilities in Python code, our first step was to extract all Python files from these projects for analysis. We developed a Python script for this purpose, and Table \ref{projects} lists the number of Python files we collected. Table \ref{methods_list} lists 26 functions from the Semgrep \citep{semgrep} database that are known to introduce command injection vulnerabilities in Python. We categorized these functions into three groups: built-in functions, functions from the \texttt{subprocess} module, and functions from the \texttt{os} module.

\begin{table}[ht]
\centering
\caption{List of Python methods prone to command injection vulnerabilities.}
\label{methods_list}
\begin{tabular}{ccc}
\hline
No.                 & \begin{tabular}[c]{@{}c@{}}Types of Python libraries and functions\end{tabular} & Methods                               \\ \hline
\multirow{2}{*}{1}  & \multirow{2}{*}{built-in function}                                                     & exec()                                \\  
                    &                                                                                        & eval()                                \\ \hline
\multirow{4}{*}{2}  & \multirow{4}{*}{subprocess module}                                                     & subprocess.call(user\_input)         \\ 
                    &                                                                                        & subprocess.run(user\_input)          \\  
                    &                                                                                        & subprocess.Popen(user\_input)        \\ 
                    &                                                                                        & subprocess.check\_output(user\_input) \\ \hline
\multirow{20}{*}{3} & \multirow{20}{*}{os module}                                                            & os.popen()                            \\  
                    &                                                                                        & os.system()                           \\  
                    &                                                                                        & os.spawnl()                           \\ 
                    &                                                                                        & os.spawnle()                          \\  
                    &                                                                                        & os.spawnlp()                          \\  
                    &                                                                                        & os.spawnlpe()                         \\  
                    &                                                                                        & os.spawnv()                           \\  
                    &                                                                                        & os.spawnve()                          \\ 
                    &                                                                                        & os.spawnvp()                          \\ 
                    &                                                                                        & os.spawnvpe()                         \\ 
                    &                                                                                        & os.posix\_spawn()                     \\ 
                    &                                                                                        & os.posix\_spawnp()                    \\  
                    &                                                                                        & os.execl()                            \\ 
                    &                                                                                        & os.execle()                           \\ 
                    &                                                                                        & os.execlp()                           \\ 
                    &                                                                                        & os.execlpe()                          \\
                    &                                                                                        & os.execv()                            \\ 
                    &                                                                                        & os.execve()                           \\
                    &                                                                                        & os.execvp()                           \\ 
                    &                                                                                        & os.execvpe()                          \\ \hline
\end{tabular}
\end{table}

To find all the candidate functions, we used a Python script to extract the Python files which contains the above methods from all the Python files in the 6 Python projects. After that, found the functions from the extracted files that contain the above methods and stored each of the found functions in separate files. Finally, we found 190 candidate functions in total and the detailed results are shown in Table \ref{projects}.

\subsection{LLM-based Vulnerability Analysis \& Security Testing Generation}
In this section, we present two important steps in our approach: using LLMs to analyze functions for potential command injection vulnerabilities and generating security test code for functions identified as potentially vulnerable.

After filtering candidate functions using the method described above, we obtained a set of functions that may contain injection vulnerabilities. However, this set is presenting uncertainty, as the presence of these methods in some functions may not necessarily lead to vulnerabilities, and there may be constraints within the functions that mitigate such risks.

To further refine this analysis, we leverage GPT-4 to generate test case to evaluate whether each candidate function is truly vulnerable to command injection. If GPT-4 determines that a function is vulnerable ("Yes"), it proceeds to generate a security test case. If GPT-4 determines that the function is safe ("No"), no command injection vulnerability is detected. 

\begin{figure}[ht]
\centering
\includegraphics[width=\textwidth]{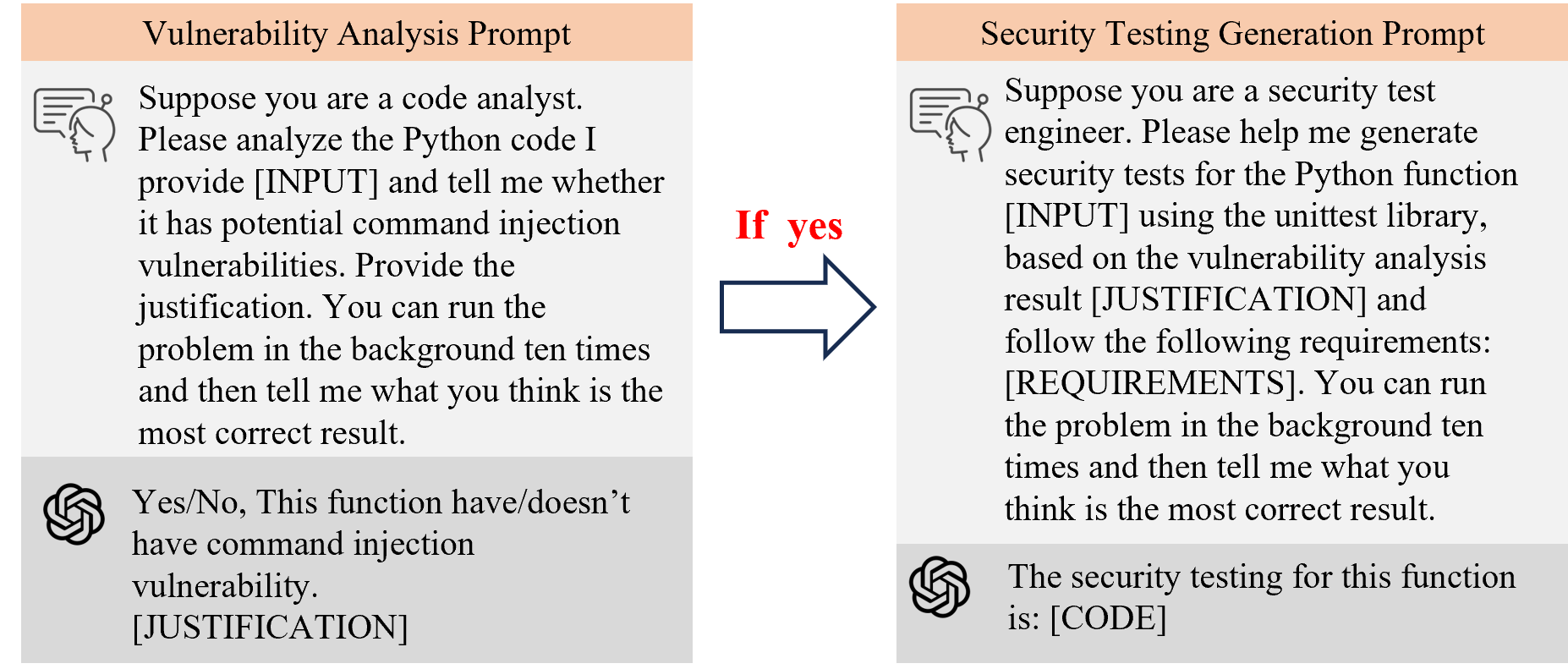}   
\caption{LLM prompt for vulnerability analysis and security testing generation.}
\label{prompt}
\end{figure}

Figure \ref{prompt} shows the prompts used for these two steps. Specifically, \colorbox{lightgray}{[INPUT]} is the Python function need to test, \colorbox{lightgray}{[JUSTIFICATION]} represents the reason provided by LLM for whether a function contains a command injection vulnerability. \colorbox{lightgray}{[CODE]} presents the security test generated by LLM. \colorbox{lightgray}{[REQUIREMENTS]} represents rules that LLM tool needs to follow in generating security test. Specifically, these rules are:
\begin{itemize}
    \item Include the source function being tested without modifying its name or content.
    \item Perform a command injection test, if there are methods in the function that would lead to a command injection attack. Generate an os command as its input to do the test. For example, create a test file and then attempt a command injection to delete it.
    \item Set the assertion section to verify if the command is executed successfully.
    \item Only generate the code; do not provide textual descriptions or suggestions.
    \item Use the unittest library, but avoid using mock modules or other simulation objects.
    \item Import any necessary libraries to run the code.
    \item Avoid redefining subprocess.call, subprocess.run, exec, or other methods in the test code.
\end{itemize}

To ensure the reliability of the GPT output, we adopted the “mimic-in-the-background” prompting method proposed in \citep{sun2023gptscan} when designing prompts for the GPT-4 model. As illustrated in Figure \ref{prompt}, the system prompts instruct the LLM to simulate answering the query in the background 10 times and then to select the response it considers most accurate. Additionally, to minimize output randomness, we set the temperature parameter for both the GPT-4 and GPT-4o models to zero.

\subsection{Validation}
\label{laststep}
\begin{figure}[ht]
\centering
\includegraphics[width=\textwidth]{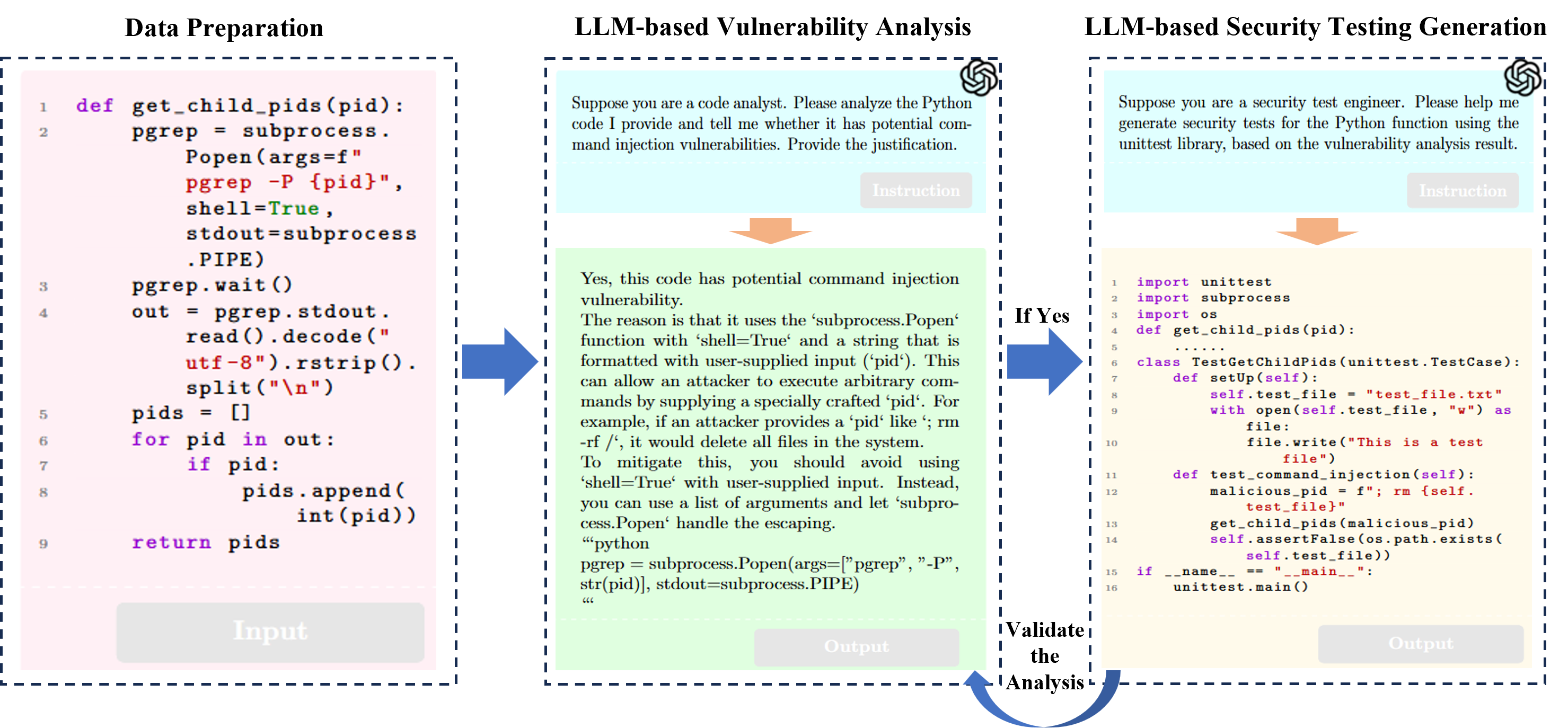}   
\caption{Example workflow of the proposed LLM-based approach.}
\label{example_LLM}
\end{figure}

Some of the test scripts generated by the GPT model require additional modifications before they can be executed. This is typically due to missing libraries, specific parameter settings needed for execution, or operating system command paths that must be adjusted to the user’s environment. To address these issues, we manually modified the test scripts to ensure compatibility with the systems used in our experiments. After making these adjustments, we were able to run the modified test scripts, allowing us to confirm which candidate functions actually contained command injection vulnerabilities. Figure \ref{example_LLM} shows a complete example of the flow of detecting command injection vulnerabilities using our approach.

%% file: evaluation.tex
\section{Evaluation} \label{sec:eval}
We conducted a detailed empirical evaluation of our proposed approach, focusing on answering three key research questions described in Section \ref{sec:rq}.

\subsection{RQ1: Effectiveness Evaluation of the Proposed Approach}

We evaluated the effectiveness of our proposed approach by applying it to 190 candidate functions drawn from six Python projects, as outlined in Section \ref{sec:data}.

Table \ref{result_list} presents the experimental results. Based on GPT-4's evaluation, 100 functions were identified as containing command injection vulnerabilities, while 90 were determined to be free from vulnerabilities.

After running the automated testing component, we observed the following outcomes:

\begin{itemize}
    \item \textbf{True Positives (TP):} 67 cases (35\%) were identified correctly, where GPT-4 determined a command injection vulnerability existed, and security testing confirmed it.
    \item \textbf{False Positives (FP):} 31 cases (16\%) were identified as false positives, where GPT-4 flagged a vulnerability, but security testing showed it did not exist.
    \item \textbf{True Negatives (TN):} 75 cases (40\%) were correctly labeled as non-vulnerable by GPT-4, confirmed by the absence of vulnerabilities in security tests.
    \item \textbf{False Negatives (FN):} 15 cases (8\%) were false negatives, where GPT-4 missed identifying a vulnerability, but manual review revealed its presence.
\end{itemize}

Additionally, there were two invalid cases (1\%) in which security tests could not run due to environmental or dependency issues.

This evaluation demonstrates the strengths and limitations of using GPT-4 for command injection vulnerability detection in Python functions, highlighting areas where LLM-based approaches may benefit from further refinement.

\begin{table*}[ht]
\caption{Detection results of command injection vulnerabilities using GPT-4.}
\label{result_list}
\centering
\begin{tabular}{ccccccc}
\hline
Project      & \begin{tabular}[c]{@{}c@{}}No. \\ of cases\end{tabular} & \begin{tabular}[c]{@{}c@{}}True \\ positive\end{tabular} & \begin{tabular}[c]{@{}c@{}}False \\ positive\end{tabular} & \begin{tabular}[c]{@{}c@{}}True \\ negative\end{tabular} & \begin{tabular}[c]{@{}c@{}}False \\ negative\end{tabular} & Invalid \\ \hline
Django       & 17                                                      & 5                                                        & 4                                                         & 7                                                        & 1                                                         & 0       \\ 
Flask        & 2                                                       & 2                                                        & 0                                                         & 0                                                        & 0                                                         & 0       \\ 
Langchain    & 13                                                      & 6                                                        & 4                                                         & 3                                                        & 0                                                         & 0       \\ 
Tensorflow   & 46                                                      & 12                                                       & 10                                                         & 23                                                       & 0                                                         & 1       \\ 
Scikit-learn & 7                                                       & 3                                                        & 2                                                         & 1                                                        & 1                                                         & 0       \\ 
Pytorch      & 105                                                     & 39                                                       & 11                                                        & 41                                                       & 13                                                         & 1       \\ 
Total        & 190                                                     & 67                                                       & 31                                                        & 75                                                       & 15                                                         & 2       \\ \hline
\end{tabular}
\end{table*}

In order to evaluate the performance of the GPT-4 model in command injection vulnerability detection more comprehensively, we selected four metrics, accuracy, precision, recall and F1 Score, for in-depth analysis. These four metrics are often used to evaluate the performance of a model\citep{SRITSE2022, WYIST2022, GYIEEE2019}, and they are calculated using the formula shown below:

\begin{equation}
\text{Accuracy} = \frac{TP + TN}{TP + TN + FP + FN}
\end{equation}

\begin{equation}
\text{Precision} = \frac{TP}{TP + FP}
\end{equation}

\begin{equation}
\text{Recall} = \frac{TP}{TP + FN}
\end{equation}

\begin{equation}
\text{F1 Score} = 2 \times \frac{\text{Precision} \times \text{Recall}}{\text{Precision} + \text{Recall}}
\end{equation}

Table \ref{metric} shows performance evaluation metrics of GPT-4 in the command injection vulnerability detection task. Of these, the accuracy was 75.5\%, the precision was 68.4\%, the recall was 81.7\% and F1 Score was 74.5\%.

\begin{table*}[ht]
\centering
\caption{Performance metrics for GPT-4 model.}
\label{metric}
\begin{tabular}{cc}
\hline
Metrics   & Results \\ \hline
Accuracy  & 75.5\%  \\ 
Precision & 68.4\%    \\ 
Recall    & 81.7\%  \\ 
F1 Score  & 74.5\%  \\ \hline
\end{tabular}
\end{table*}

\begin{tcolorbox}[colback=lightgray!20, colframe=gray!10, boxrule=0pt, sharp corners]

\textbf{Answer for RQ1:} 
The GPT-4 model analyzed 190 functions with possible command injection vulnerabilities, showing its ability to detect command injection vulnerabilities with a accuracy of 75.5\%, a precision of 68.4\%, a recall of 81.7\% and an F1 score of 74.5\%.
\end{tcolorbox}

\subsection{RQ2: Completeness}

Table \ref{result_list} identified 15 false-negative cases, which were actual command injection vulnerabilities in the functions that \texttt{GPT-4} did not detect. This outcome raises an important concern about the types of vulnerabilities that might be challenging for LLM-based methods to identify. As such, in this section we further analyzed these 14 cases to identify characteristics of vulnerabilities that GPT-4 might miss.

Table \ref{FN_list} provides a detailed list of these false-negative cases, which include one case each from the Scikit-learn and Django projects and 13 cases from the PyTorch project. Of these fourteen cases, 10 are command injection vulnerabilities related to the \texttt{subprocess} module methods, such as \texttt{subprocess.run()}, \\
\texttt{subprocess.Popen()}, and \texttt{subprocess.check\_output()}. The remaining 5 cases involve vulnerabilities in the use of the \texttt{eval()} and \texttt{exec()} function.

To understand the problem, we examined the code for each case. In 9 of the 10 cases related to the \texttt{subprocess} module, the code followed a structure similar to that shown in Listing \ref{ignoretype}. Specifically, the \texttt{args} parameter passed to \texttt{subprocess.run()} or similar methods was a list of strings, a format that GPT-4 generally disregarded as a potential command injection vulnerability. The remaining case from the Scikit-learn project involved a vulnerability where a global variable, modifiable by an attacker, was passed to the \texttt{subprocess.check\_output()} method, allowing for the injection of malicious commands.

In the 5 cases from the PyTorch project that involved the \texttt{eval()} and \texttt{exec()} function, the vulnerabilities were introduced by passing parameters to \texttt{eval()} and \texttt{exec()} that could be externally modified.

These findings reveal that GPT-4 may miss certain command injection vulnerabilities, particularly when specific code structures or parameter types obscure the risk. This underscores the need for further refinement in LLM-based vulnerability detection methods to improve accuracy in identifying subtle or complex injection risks.

\begin{lstlisting}[caption={Types of command injection vulnerabilities that would be ignored by the GPT-4 model.},label=ignoretype]
def candidate_function(args: List[str]):
    return  subprocess.run(args,
        capture_output=True,
        check=True,
        )
\end{lstlisting}

\begin{table}[ht]
\caption{Details of false negative cases and associated command injection vulnerabilities.}
\label{FN_list}
\centering
\begin{tabular}{cccc}
\hline
Project Name              & \begin{tabular}[c]{@{}c@{}}Case\\ No.\end{tabular} & \begin{tabular}[c]{@{}c@{}}Line of\\ Code\end{tabular} & \begin{tabular}[c]{@{}c@{}}Method that \\ may cause\\ vulnerability\end{tabular} \\ \hline
Scikit-learn              & 5                                                  & 17                                                     & subprocess.check\_output()                                                       \\ \hline
Django                    & 6                                                  & 32                                                     & subprocess.run()                                                                 \\ \hline
\multirow{12}{*}{Pytorch} & 4                                                  & 67                                                     & eval()                                                                           \\ 
                          & 4.1                                                & 49                                                     & eval()                                                                           \\  
                          & 17                                                 & 69                                                     & subprocess.run()                                                                 \\ 
                          & 23                                                 & 13                                                     & subprocess.run()                                                                 \\ 
                          & 24                                                 & 30                                                     & exec()                                                                          \\ 
                          & 28                                                 & 13                                                     & subprocess.Popen()                                                               \\ 
                          & 30                                                 & 13                                                     & subprocess.run()                                                                 \\ 
                          & 31                                                 & 72                                                     & eval()                                                                           \\ 
                          & 75                                                 & 13                                                     & subprocess.run()                                                                 \\ 
                          & 76                                                 & 33                                                     & eval()                                                                           \\ 
                          & 79                                                 & 22                                                     & subprocess.run()                                                                 \\
                          & 80                                                 & 13                                                     & subprocess.run()                                                                 \\  
                          & 84                                                 & 35                                                     & subprocess.run()                                                                 \\ \hline
\end{tabular}
\end{table}

\begin{tcolorbox}[colback=lightgray!20, colframe=gray!10, boxrule=0pt, sharp corners]
\textbf{Answer for RQ2:} The LLM-based approach based on \texttt{GPT-4} model demonstrates limitations in accurately detecting command injection vulnerabilities within the \texttt{subprocess} family when list-type parameters are used (e.g., \texttt{subprocess.run(args)} or \texttt{subprocess.Popen(args)}, where \texttt{args} is a list of strings). Although vulnerabilities missed by GPT-4 due to global variables or externally modifiable parameters are relatively few, these potential security risks require attention to prevent them from being overlooked in real-world applications.
\end{tcolorbox}

\subsection{RQ3: Running Cost}

In RQ3, we evaluated the runtime and financial costs associated with performing all experiments. Table \ref{Cost_list} provides details on the number of cases, lines of code, and time spent analyzing these cases with using \texttt{GPT-4}.

Our experiment has analyzed $190$ candidate functions, comprising a total of $5239$ lines of code. Among these, $100$ functions were identified as containing command injection vulnerabilities, for which GPT-4 analyzed $2117$ lines of code and generated corresponding security tests. The total time required for this analysis was $3629.8$ seconds. For the remaining 90 functions, which were determined to be free from command injection vulnerabilities, GPT-4 analyzed $3122$ lines of code, taking a total of $950.99$ seconds.

\begin{table*}[ht]
\centering
\caption{Time costs of performing experiments with GPT-4.}
\label{Cost_list}
\begin{tabular}{|c|c|c|c|}
\hline
Analysis Result & No. of case & Line of code & Time    \\ \hline
Yes             & 100         & 2117         & 3629.8  \\ \hline
No              & 90          & 3122         & 950.99  \\ \hline
Total           & 190         & 5239         & 4580.79 \\ \hline
\end{tabular}
\end{table*}

\begin{tcolorbox}[colback=lightgray!20, colframe=gray!10, boxrule=0pt, sharp corners]
\textbf{Answer for RQ3:} We used the GPT-4 model to analyze a total of 190 candidate functions with a total of 5239 lines of code, of which 100 candidates were determined to have command injection vulnerabilities, and GPT generated the corresponding security tests for these 100 cases. For all experiments, the total time spent was 4580.79 seconds and the financial overhead was \$14.19.
\end{tcolorbox}

\subsection{RQ4: Comparison of Performance between Different LLM tools }

In this section, we compare the performance of our approach when integrating four popular Large Language Models (LLMs)—Claude 3.5 Sonnet, GPT-4o, GPT-4 and DeepSeek-R1—in tasks of command injection vulnerability detection and automated security test generation. By examining each model’s accuracy in vulnerability detection and the percentage of generated security tests that can be executed without modifications, we gain insights into their capabilities and reliability, helping us identify the most suitable model for our tasks in vulnerability detection and test case generation.

\subsubsection{Command Injection Vulnerability Detection Capability}

Tables \ref{claude_result_list}, \ref{gpt4o_result_list} and  \ref{deepseek_result_list} present the results of analyzing 190 candidate functions using Claude 3.5 Sonnet, GPT-4o and DeepSeek-R1. According to Claude's assessment, 156 of the 190 cases were identified as vulnerable to command injection, while 34 were considered safe. Among these, 73 were true positives, 81 were false positives, 25 were true negatives, 9 were false negatives, and 2 were invalid cases. For GPT-4o, it identified 131 cases as containing command injection vulnerabilities and 59 as safe, resulting in 66 true positives, 63 false positives, 43 true negatives, 16 false negatives, and 2 invalid cases. For DeepSeek-R1, 84 of the 190 cases were identified as vulnerable to command injection, while 106 were considered safe. Among these, 56 were true positives, 26 were false positives, 80 were true negatives, 26 were false negatives, and 2 were invalid cases.

\begin{figure}[ht]
  \centering
  \includegraphics[width=\textwidth]{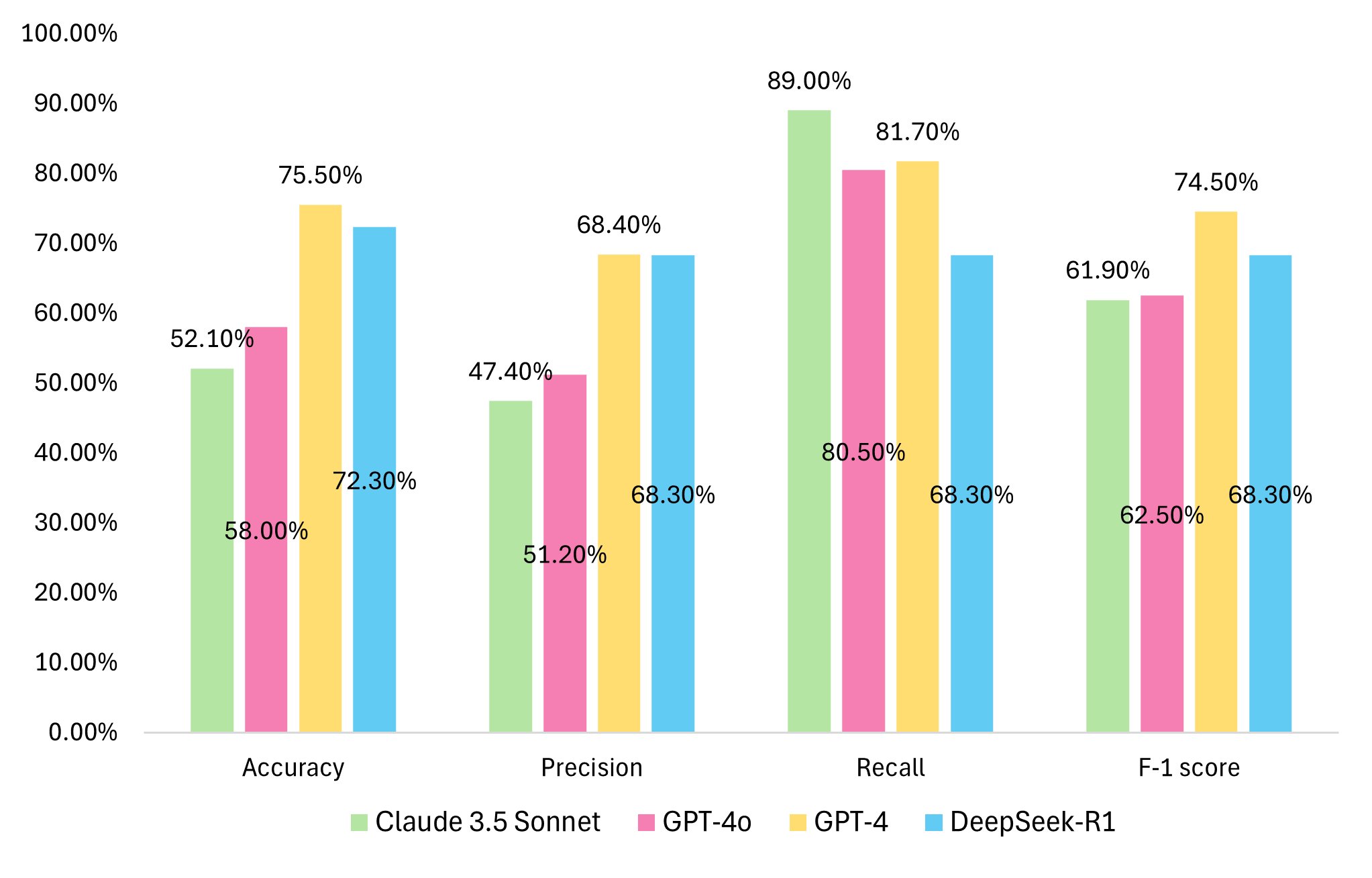}
  \caption{The comparison results of the four LLMs in command injection vulnerabilities detection performance.}
  \label{vulnerability_detection_compare}
\end{figure}

Figure \ref{vulnerability_detection_compare} illustrates the performance metrics for the four models in command injection vulnerability detection.In terms of accuracy and precision, DeepSeek-R1 and GPT‑4 perform nearly identically. By contrast, GPT‑4o and Claude 3.5 Sonnet score significantly lower on both metrics — hovering around 50\%, roughly 20 percentage points below the results achieved by GPT‑4 and DeepSeek‑R1.
For the F1 score, the GPT-4 has the highest value of 74.5\%. In comparison, GPT-4o and Claude 3.5 Sonnet had F1 scores of 62.5\% and 61.9\% respectively, which were about 10 percentage points lower than GPT-4. DeepSeek-R1's F1 score of 68.3\% is about 6 percentage points lower than GPT-4. 
However, the recall rates for GPT-4o (80.5\%) and GPT-4 (81.7\%) were comparable, while Claude 3.5 Sonnet achieves the highest recall at 89\%. DeepSeek‑R1 records the lowest recall rate, at 68.3\%.

These comparisons indicate that GPT-4 outperforms GPT-4o, Claude 3.5 Sonnet and DeepSeek‑R1 in terms of overall accuracy and F1 score, making it a more reliable choice for command injection vulnerability detection. As such, GPT-4 is recommended as the preferred model for our task.

\begin{table*}[ht]
\caption{Detection results of command injection vulnerabilities using Claude 3.5 Sonnet.}
\label{claude_result_list}
\centering
\begin{tabular}{ccccccc}
\hline
Project      & \begin{tabular}[c]{@{}c@{}}No. \\ of cases\end{tabular} & \begin{tabular}[c]{@{}c@{}}True\\ positive\end{tabular} & \begin{tabular}[c]{@{}c@{}}False\\ positive\end{tabular} & \begin{tabular}[c]{@{}c@{}}True\\ negative\end{tabular} & \begin{tabular}[c]{@{}c@{}}False\\ negative\end{tabular} & Invalid \\ \hline
Django       & 17                                                      & 6                                                       & 8                                                       & 3                                                       & 0                                                        & 0       \\ 
Flask        & 2                                                       & 2                                                       & 0                                                        & 0                                                       & 0                                                        & 0       \\ 
Langchain    & 13                                                      & 6                                                       & 7                                                        & 0                                                       & 0                                                        & 0       \\ 
Tensorflow   & 46                                                      & 12                                                      & 25                                                       & 8                                                       & 0                                                        & 1       \\ 
Scikit-learn & 7                                                       & 4                                                       & 2                                                       & 1                                                       & 0                                                        & 0       \\ 
Pytorch      & 105                                                     & 43                                                      & 39                                                       & 13                                                       & 9                                                        & 1       \\ 
Total        & 190                                                     & 73                                                      & 81                                                      & 25                                                       & 9                                                        & 2       \\ \hline
\end{tabular}
\end{table*}

\begin{table*}[ht]
\caption{Detection results of command injection vulnerabilities using GPT-4o.}
\label{gpt4o_result_list}
\centering
\begin{tabular}{ccccccc}
\hline
Project      & \begin{tabular}[c]{@{}c@{}}No. \\ of cases\end{tabular} & \begin{tabular}[c]{@{}c@{}}True\\ positive\end{tabular} & \begin{tabular}[c]{@{}c@{}}False\\ positive\end{tabular} & \begin{tabular}[c]{@{}c@{}}True\\ negative\end{tabular} & \begin{tabular}[c]{@{}c@{}}False\\ negative\end{tabular} & Invalid \\ \hline
Django       & 17                                                      & 4                                                       & 2                                                       & 9                                                       & 2                                                        & 0       \\ 
Flask        & 2                                                       & 2                                                       & 0                                                        & 0                                                       & 0                                                        & 0       \\ 
Langchain    & 13                                                      & 6                                                       & 6                                                        & 1                                                       & 0                                                        & 0       \\ 
Tensorflow   & 46                                                      & 11                                                      & 25                                                       & 8                                                       & 1                                                       & 1       \\ 
Scikit-learn & 7                                                       & 3                                                       & 1                                                       & 2                                                       & 1                                                        & 0       \\
Pytorch      & 105                                                     & 40                                                      & 29                                                       & 23                                                       & 12                                                        & 1       \\ 
Total        & 190                                                     & 66                                                      & 63                                                      & 43                                                       & 16                                                        & 2       \\ \hline
\end{tabular}
\end{table*}

\begin{table*}[ht]
\caption{Detection results of command injection vulnerabilities using DeepSeek-R1.}
\label{deepseek_result_list}
\centering
\begin{tabular}{ccccccc}
\hline
Project      & \begin{tabular}[c]{@{}c@{}}No. \\ of cases\end{tabular} & \begin{tabular}[c]{@{}c@{}}True\\ positive\end{tabular} & \begin{tabular}[c]{@{}c@{}}False\\ positive\end{tabular} & \begin{tabular}[c]{@{}c@{}}True\\ negative\end{tabular} & \begin{tabular}[c]{@{}c@{}}False\\ negative\end{tabular} & Invalid \\ \hline
Django       & 17                                                      & 6                                                       & 2                                                       & 9                                                       & 0                                                        & 0       \\ 
Flask        & 2                                                       & 2                                                       & 0                                                        & 0                                                       & 0                                                        & 0       \\ 
Langchain    & 13                                                      & 6                                                       & 2                                                        & 5                                                       & 0                                                        & 0       \\ 
Tensorflow   & 46                                                      & 9                                                      & 5                                                       & 28                                                       & 3                                                       & 1       \\ 
Scikit-learn & 7                                                       & 2                                                       & 1                                                       & 2                                                       & 2                                                        & 0       \\ 
Pytorch      & 105                                                     & 31                                                      & 16                                                       & 36                                                       & 21                                                        & 1       \\ 
Total        & 190                                                     & 56                                                      & 26                                                      & 80                                                       & 26                                                        & 2       \\ \hline
\end{tabular}
\end{table*}

\subsubsection{The security test generation capability}
\begin{figure}[ht]
  \centering
  \includegraphics[width=\textwidth]{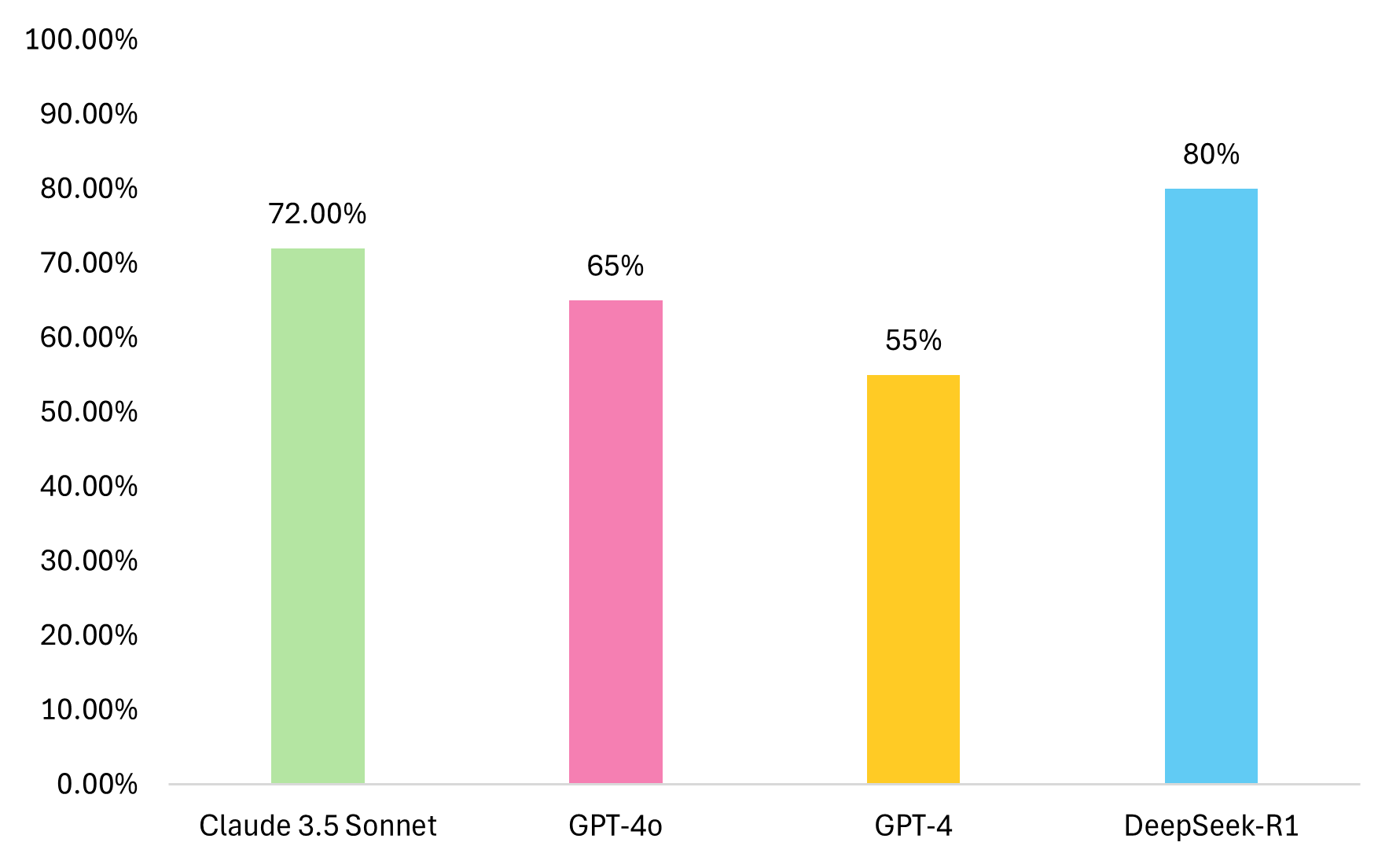}
  \caption{The comparison results of the four LLMs in security test generation performance.}
  \label{test_generation_compare}
\end{figure}

Our experiment has identified that GPT-4 determined that 100 out of 190 cases had command injection vulnerabilities and generated security tests for them, among which only 55 security tests can be run directly, which indicates that GPT-4 had deficiencies in security test generation. For further comparison, we used Claude 3.5 Sonnet, GPT-4o and DeepSeek-R1 to generate security tests for these 100 cases. Figure \ref{test_generation_compare} showed the comparison results of the four LLMs in security test generation performance. For Claude 3.5 Sonnet, the number of security tests that can be directly run is 72. The number of security tests that can be directly run for GPT-4o is 65. And for DeepSeek-R1, the number of security tests that can be directly run is 80. Among the four LLMs, DeepSeek-R1 performed best in generating security tests. It is the better choice in test generation task.

\begin{tcolorbox}[colback=lightgray!20, colframe=gray!10, boxrule=0pt, sharp corners]
\textbf{Answer for RQ4:} Among the four large language models, GPT-4 showed better results in command injection vulnerability detection, while DeepSeek-R1 showed better results in security test generation. We can choose different large language models for different tasks to get better results.
\end{tcolorbox}

\subsection{RQ5: Comparison with Bandit}
\begin{figure}[ht]
    \centering
    \begin{subfigure}{0.33\textwidth}        
        \includegraphics[width=\textwidth]{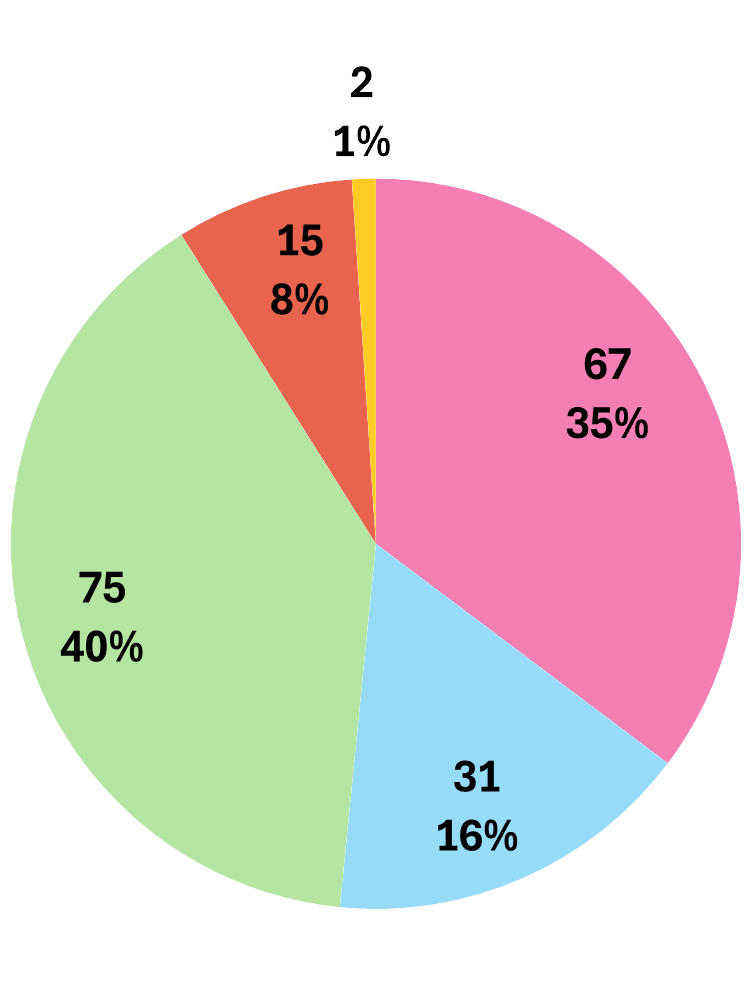}      
        \captionsetup{justification=centering}
        \caption{Distribution of GPT-4}
        \label{figureGPT4_result}
    \end{subfigure}
    \begin{subfigure}{0.2\textwidth}        
        \includegraphics[width=\textwidth]{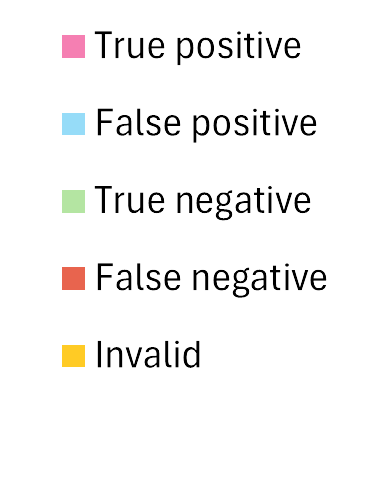}      
    \end{subfigure}
    \begin{subfigure}{0.35\textwidth}
        \includegraphics[width=\textwidth]{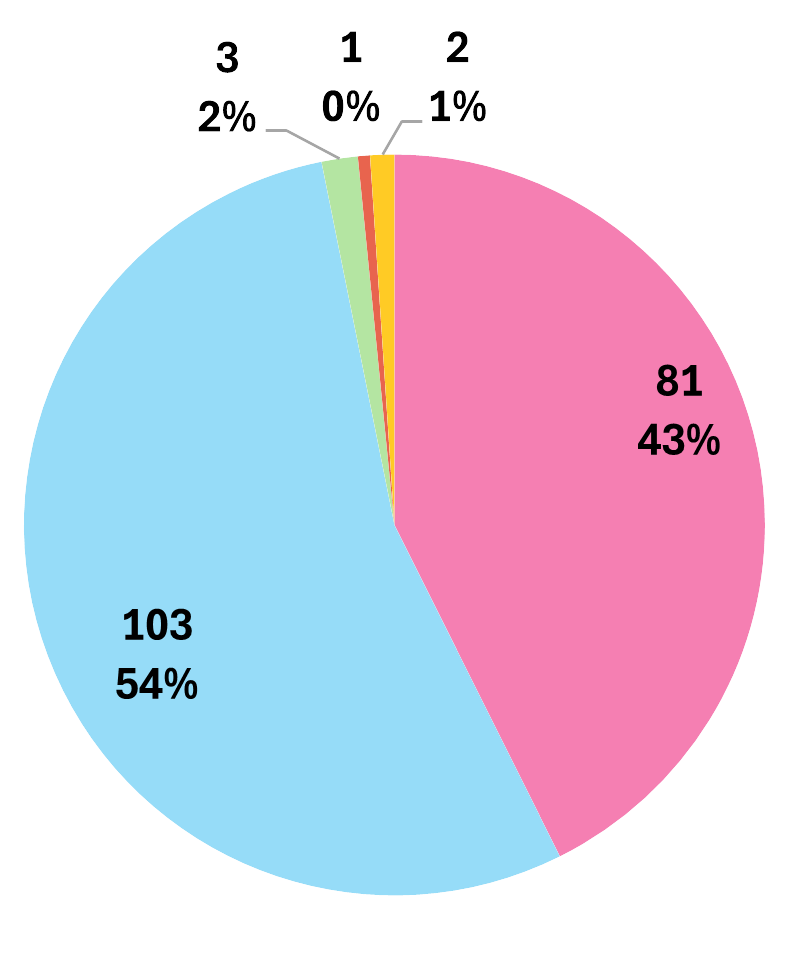} 
        \captionsetup{justification=centering}
        \caption{Distribution of Bandit}
        \label{bandit_result}
    \end{subfigure}
    \caption{The distribution of detection results for GPT-4 and Bandit.}
    \label{two methods experiment results}
\end{figure}

We compared our approach with an existing Python security detection tool, Bandit, focusing on command injection vulnerability analysis. Given that the GPT-4 model demonstrated the highest performance among the four large language models tested, we specifically compared GPT-4's results with those of Bandit. Table \ref{bandit_result_list} show Bandit's performance for detecting vulnerabilities in 190 candidate functions. Bandit identified 186 functions as containing command injection vulnerabilities and 4 as safe. Upon verification, 81 cases were true positives (43\%), 103 were false positives (54\%), 3 were true negatives (1.5\%), and 1 was a false negative (0.5\%). Additionally, 2 cases (1\%) were invalid due to environmental or dependency issues. Figure \ref{two methods experiment results} shows the distribution of detection results for GPT-4 and Bandit. We can find that, in contrast, Bandit has a much higher false positive rate than GPT-4.

\begin{table*}[ht]
\caption{Detection results of command injection vulnerabilities using Bandit.}
\label{bandit_result_list}
\centering
\begin{tabular}{ccccccc}
\hline
Project      & \begin{tabular}[c]{@{}c@{}}No. \\ of cases\end{tabular} & \begin{tabular}[c]{@{}c@{}}True\\ positive\end{tabular} & \begin{tabular}[c]{@{}c@{}}False\\ positive\end{tabular} & \begin{tabular}[c]{@{}c@{}}True\\ negative\end{tabular} & \begin{tabular}[c]{@{}c@{}}False\\ negative\end{tabular} & Invalid \\ \hline
Django       & 17                                                      & 6                                                       & 10                                                       & 1                                                       & 0                                                        & 0       \\ 
Flask        & 2                                                       & 2                                                       & 0                                                        & 0                                                       & 0                                                        & 0       \\ 
Langchain    & 13                                                      & 6                                                       & 6                                                        & 1                                                       & 0                                                        & 0       \\ 
Tensorflow   & 46                                                      & 12                                                      & 32                                                       & 1                                                       & 0                                                        & 1       \\ 
Scikit-learn & 7                                                       & 4                                                       & 3                                                        & 0                                                       & 0                                                        & 0       \\ 
Pytorch      & 105                                                     & 51                                                      & 52                                                       & 0                                                       & 1                                                        & 1       \\ 
Total        & 190                                                     & 81                                                      & 103                                                      & 3                                                       & 1                                                        & 2       \\ \hline
\end{tabular}
\end{table*}

\begin{figure}[ht]
  \centering
  \includegraphics[width=\textwidth]{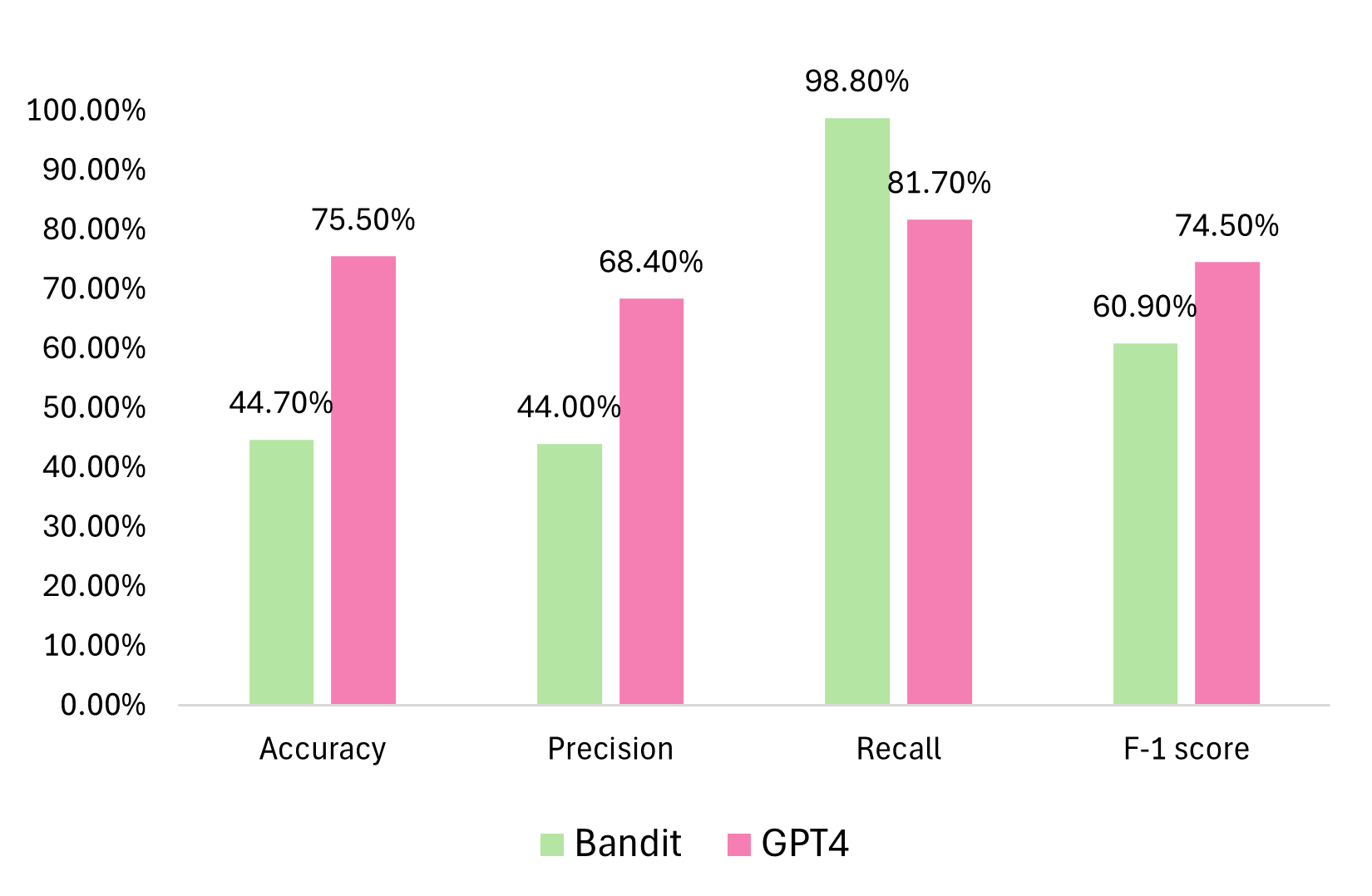}
  \caption{Performance comparison of GPT-4 and Bandit in detecting command injection vulnerabilities.}
  \label{compare_result}
\end{figure}

 Figure \ref{compare_result} presented the performance comparison of using Bandit and GPT-4 in detecting command injection vulnerabilities. Compared to Bandit, our GPT-4 based method performs better in detecting command injection vulnerabilities with improved accuracy of 30.8\%, precision of 24.4\%, and F1 score of 13.6\%. The only shortcoming is that the recall is reduced by 17.1\%. The recall value of Bandit is higher than GPT-4, indicating that Bandit has stronger sensitivity to positive cases and can effectively reduce the number of false negatives. However, we observed that Bandit has a high number of false positives, which significantly lowers its precision and accuracy, with values of 44\% and 44.7\%, respectively. It is worth mentioning that our method is also able to identify two vulnerabilities that Bandit cannot detect. These results indicate that the large language model performs better in command injection vulnerability detection compared to static vulnerability detection tools because of its stronger contextual analysis capability.

\begin{tcolorbox}[colback=lightgray!20, colframe=gray!10, boxrule=0pt, sharp corners]
\textbf{Answer to RQ5:} Our LLM-based approach outperformed Bandit by reducing false positive and false negative rates, thereby significantly improving accuracy, precision, and F1 scores. Additionally, our method successfully identified vulnerabilities that Bandit failed to detect, highlighting the enhanced detection capability of our approach.
\end{tcolorbox}

%% file: related_work.tex
\section{Related Work}
\subsection{Vulnerability detection}
We categorize research related to vulnerability detection into three main types: static or dynamic analysis methods, machine learning-based methods, and large language models-based methods. 

\textbf{Static analysis and dynamic analysis approaches}
Static analysis and dynamic analysis are the classical methods to detect vulnerability. Static analysis is divided into two categories: graph-based static analysis \citep{10.1007/978-3-540-89862-7_1} and static analysis with data modeling \citep{6956589}. For example, Song et al. \citep{10.1007/978-3-540-89862-7_1} proposed an approach called BitBlaze, where a static analysis component called Vine can detect vulnerabilities using CFG, DFG, and weakest precondition calculations. 

Dynamic analysis mainly consists of detection techniques such as dynamic taint analysis \citep{cai2016sworddta}, fuzzing testing \citep{10179317}, and symbolic execution \citep{li2013software}. For example, Trickel et al. \citep{10179317} proposed a novel Web vulnerability discovery framework based on grey-box coverage-guided fuzzing called Witcher. It implements the concept of fault escalation to detect SQL and command injection vulnerabilities in web applications.
\textbf{ML-based approaches:}
Compared to static and dynamic analysis, machine learning based approaches handle large scale data as well as reduce false positives in vulnerability detection.
Both Guo \citep{10.1007/978-3-030-41579-2_12} and Laura \citep{Wartschinski_2022} proposed vulnerability detection methods based on Long Short-Term Memory, where Guo's method is used to detect vulnerabilities in PHP code, while Laura's method is used to detect vulnerabilities in Python code. After evaluation, the accuracy and F1 scores of both methods are more than 80\%.
Stanislav \citep{8835902} used a CNN-based approach to detect code injection in web application and uses data preprocessing techniques to address the large training requirements typically associated with such networks, reducing the time required to configure and train CNNs.

\textbf{LLMs-based approaches:}
With the development of large language models, more and more people are using this technique to analyze code \citep{ouyang2023llm, dai2023llm} and detect vulnerabilities \citep{hu2023large, liu2023harnessing, sun2023gptscan}. Liu et al. \citep{liu2023harnessing} proposed a static binary taint analysis method supported by LLM that automates the taint analysis process and outperforms state-of-the-art methods in terms of efficiency and effectiveness in identifying new errors in realistic firmware. And Sun et al. \citep{sun2023gptscan} combined the GPT model with static analysis to detect logical vulnerabilities in smart contracts with high accuracy. Their approach has been a great inspiration for our research.

\subsection{LLMs-assisted techniques in unit test generation} 
Recently, as test generation has been moving towards automation, more and more people have been trying to generate tests using the emerging technique of large language models \citep{10329992, tang2023chatgpt, 10.1145/3624032.3624035, yuan2023manual, xie2023chatunitest, zhang2023does}. Most of them use the GPT family of models (GPT-3.5 or GPT-4) or its variant GPT to generate tests. For example, Max et al. \citep{10329992} propose a method for generating adaptive unit tests using a large language model called TESTPILOT. This method is based on OpenAI's GPT-3.5-turbo model and was evaluated on 25 npm packages. The statement coverage and branch coverage of the generated tests are better than the state-of-the-art feedback-directed JavaScript test generation technique. Zhang et al. \citep{zhang2023does} evaluated the effectiveness of GPT in generating security tests for Java applications. They used GPT-4 to generate security tests for 55 applications, 40 of which could be compiled and successfully demonstrated 24 attacks. The results are better than two of the most advanced security test generators (TRANSFER and SIEGE).

%% file: threats_to_validity.tex
\section{Threats to Validity}
\label{threats}
In this section, we focus on discussing the potential internal and external threats to the validity.

\textbf{Internal threats:}
Our study adopted a validation strategy, i.e., validating the results of LLM's analysis of the code by running the security tests generated by it. We checked all the generated security tests and manually modified the code that could not be run. Finally, in 100 security tests, 2 tests could not run due to code dependency issues and the running environment issues. The percentage is 2\%. They slightly affected the accuracy of our assessment of LLMs' ability to detect vulnerabilities. 

\textbf{External threats:}
Our study only uses four large language model(GPT-4, GPT-4o, Claude 3.5 Sonnet and DeepSeek-R1) and tests our approach in six Python projects. So there are some limitations to our research results. Future research will explore more large language models (e.g., DeBERTa, LLaMA, etc.) and add more Python projects to our dataset to further validate and generalize our findings.

%% file: conclusion.tex
\section{Conclusion and future work}

This study demonstrates the potential of Large Language Models (LLMs) as an effective approach for detecting command injection vulnerabilities within Python’s widely used open-source libraries. Through a comprehensive analysis of over 13,000 Python files from six high-profile projects, we evaluated the strengths and limitations of LLMs in identifying security vulnerabilities that threaten system integrity and data privacy. While traditional tools like Bandit are valuable, our results show that LLMs provide a complementary advantage by analyzing fragmented and non-compilable code and detecting complex vulnerability patterns that existing tools may miss. Additionally, the ability of LLMs to generate security tests adds a useful layer of verification, potentially enhancing the accuracy of vulnerability assessments.

Our comparative analysis of different LLM tools highlights that models like GPT-4 offer adaptability for security applications, although some challenges in vulnerability detection persist. By identifying the types of vulnerabilities that LLMs might miss, our research lays a foundation for enhancing these models and guiding future developments in automated security analysis. The dataset we built, sourced from six extensively used GitHub projects, aims to support further research, establishing a benchmark for LLM-driven vulnerability detection. With these advancements, developers and security researchers can leverage LLMs to improve security practices, moving toward a more resilient open-source ecosystem.

Future work can explore refining LLMs for greater accuracy in vulnerability detection, particularly focusing on areas where current models fall short, such as complex nested code structures. Investigating hybrid models that combine LLMs with traditional static analysis tools could also yield improved detection capabilities. Additionally, expanding the dataset to include a broader range of security vulnerabilities would create a more robust benchmark for future studies.

%% file: appendix.tex
\begin{table}[ht]
\resizebox{1\textwidth}{!}{
                    & Yes                                                                      & Yes                                                             \\ 
                          & Case 27                                                 & 43                                                      & subprocess.check\_output()                                                              & No                                                      & 12.52                                                               & N/A                                                                          & N/A                                                                                          & N/A                                                                      & No                                                              \\ 
                          & Case 28                                                 & 13                                                      & subprocess.Popen()                                                                      & No                                                      & 11.36                                                               & N/A                                                                          & N/A                                                                                          & N/A                                                                      & Yes                                                             \\ 
                          & Case 28.1                                               & 10                                                      & subprocess.Popen()                                                                      & Yes                                                     & 26.75                                                               & Yes                                                                          & N/A                                                                                          & N/A                                                                      & Yes                                                             \\ 
                          & Case 29                                                 & 9                                                       & eval()                                                                                  & Yes                                                     & 47                                                                  & Yes                                                                          & N/A                                                                                          & N/A                                                                      & No                                                              \\
                          & Case 30                                                 & 13                                                      & subprocess.run()                                                                        & No                                                      & 11.47                                                               & N/A                                                                          & N/A                                                                                          & N/A                                                                      & Yes                                                             \\ 
                          & Case 31                                                 & 72                                                      & eval()                                                                                  & No                                                      & 8.34                                                                & N/A                                                                          & N/A                                                                                          & N/A                                                                      & Yes                                                              \\ 
                          & Case 32                                                 & 38                                                      & exec()                                                                                  & No                                                      & 4.94                                                                & N/A                                                                          & N/A                                                                                          & N/A                                                                      & No                                                              \\  
                          & Case 33                                                 & 6                                                       & subprocess.check\_output()                                                              & No                                                      & 7.84                                                                & N/A                                                                          & N/A                                                                                          & N/A                                                                      & No                                                              \\ 
                          & Case 34                                                 & 16                                                      & subprocess.run()                                                                        & Yes                                                     & 30.42                                                               & Yes                                                                          & N/A                                                                                          & N/A                                                                      & Yes                                                             \\ 
                          & Case 35                                                 & 70                                                      & subprocess.check\_output()                                                              & No                                                      & 7.13                                                                & N/A                                                                          & N/A                                                                                          & N/A                                                                      & No                                                              \\ 
                          & Case 36                                                 & 20                                                      & subprocess.run()                                                                        & Yes                                                     & 44.68                                                               & Yes                                                                          & N/A                                                                                          & N/A                                                                      & Yes                                                             \\ 
                          & Case 37                                                 & 66                                                      & exec()                                                                                  & No                                                     & 14.83                                                              & N/A                                                                         & N/A & N/A                                                                      & No                                                              \\  
                          & Case 38                                                 & 10                                                      & subprocess.check\_output()                                                              & Yes                                                     & 41.77                                                               & Yes                                                                          & N/A                                                                                          & N/A                                                                      & No                                                              \\ 
                          & Case 39                                                 & 23                                                      & exec()                                                                                  & No                                                      & 11.9                                                                & N/A                                                                          & N/A                                                                                          & N/A                                                                      & No                                                              \\ 
                          & Case 40                                                 & 15                                                      & subprocess.check\_output()                                                              & Yes                                                     & 45.68                                                               & Yes                                                                          & N/A                                                                                          & N/A                                                                      & No                                                              \\ 
                          & Case 41                                                 & 83                                                      & exec()                                                                                  & No                                                      & 13.22                                                               & N/A                                                                          & N/A                                                                                          & N/A                                                                      & No                                                              \\ 
                          & Case 42                                                 & 53                                                      & subprocess.run()                                                                        & No                                                      & 8.14                                                                & N/A                                                                          & N/A                                                                                          & N/A                                                                      & No                                                              \\
                          & Case 43                                                 & 40                                                      & exec()                                                                                  & No                                                      & 8.18                                                                & N/A                                                                          & N/A                                                                                          & N/A                                                                      & No                                                              \\ 
                          & Case 44                                                 & 16                                                      & exec()                                                                                  & No                                                      & 22.5                                                                & N/A                                                                          & N/A                                                                                          & N/A                                                                      & No                                                              \\ 
                          & Case 45                                                 & 44                                                      & subprocess.Popen()                                                                      & No                                                      & 7.97                                                                & N/A                                                                          & N/A                                                                                          & N/A                                                                      & No                                                              \\ 
                          & Case 46                                                 & 14                                                      & subprocess.run()                                                                        & Yes                                                     & 31.65                                                               & Yes                                                                          & N/A                                                                                          & N/A                                                                      & Yes                                                             \\ 
                          & Case 47                                                 & 17                                                      & subprocess.check\_output()                                                              & Yes                                                     & 48.1                                                                & Yes                                                                          & N/A                                                                                          & N/A                                                                      & No                                                              \\
                          & Case 48                                                 & 89                                                      & subprocess.check\_output()                                                              & No                                                      & 15.21                                                               & N/A                                                                          & N/A                                                                                          & N/A                                                                      & No                                                              \\  
                          & Case 49                                                 & 7                                                       & exec()                                                                                  & Yes                                                     & 36.45                                                               & Yes                                                                          & N/A                                                                                          & N/A                                                                      & Yes                                                             \\ 
                          & Case 50                                                 & 23                                                      & subprocess.run()                                                                        & Yes                                                     & 45.36                                                               & Yes                                                                          & N/A                                                                                          & N/A                                                                      & No                                                                                                                                                                                       \\ \hline
\end{tabular}
}
\end{table}

\clearpage
\begin{table}[p]
\centering
\resizebox{1\textwidth}{!}{
             & Yes                                                                      & Yes                                                             \\ 
                          & Case 83                                                 & 8                                                       & subprocess.run()                                                                  & Yes                                                     & 13.97                                                               & Yes                                                                          & N/A                                                                                           & N/A                                                                      & Yes                                                             \\ 
                          & Case 84                                                 & 35                                                      & subprocess.run()                                                                  & No                                                      & 6.28                                                                & N/A                                                                          & N/A                                                                                           & N/A                                                                      & Yes                                                             \\ 
                          & Case 85                                                 & 4                                                       & subprocess.run()                                                                  & Yes                                                     & 10.27                                                               & Yes                                                                          & N/A                                                                                           & N/A                                                                      & Yes                                                             \\ 
                          & Case 86                                                 & 31                                                      & subprocess.Popen()                                                                & No                                                      & 5.82                                                                & N/A                                                                          & N/A                                                                                           & N/A                                                                      & No                                                              \\ 
                          & Case 87                                                 & 14                                                      & subprocess.Popen()                                                                & Yes                                                     & 13.77                                                               & Yes                                                                          & N/A                                                                                           & N/A                                                                      & Yes                                                             \\
                          & Case 88                                                 & 14                                                      & subprocess.Popen()                                                                & Yes                                                     & 11.55                                                               & Yes                                                                          & N/A                                                                                           & N/A                                                                      & Yes                                                             \\ 
                          & Case 89                                                 & 40                                                      & subprocess.Popen()                                                                & Yes                                                     & 17.56                                                               & Yes                                                                          & N/A                                                                                           & N/A                                                                      & No                                                              \\ 
                          & Case 90                                                 & 29                                                      & subprocess.Popen()                                                                & No                                                      & 8.55                                                                & N/A                                                                          & N/A                                                                                           & N/A                                                                      & No                                                              \\ 
                          & Case 91                                                 & 3                                                       & subprocess.Popen()                                                                & No                                                      & 4.37                                                                & N/A                                                                          & N/A                                                                                           & N/A                                                                      & No                                                              \\ 
                          & Case 92                                                 & 11                                                      & subprocess.Popen()                                                                & Yes                                                     & 17.23                                                               & Yes                                                                          & N/A                                                                                           & N/A                                                                      & Yes                                                             \\
                          & Case 93                                                 & 20                                                      & subprocess.Popen()                                                                & No                                                      & 3.81                                                                & N/A                                                                          & N/A                                                                                           & N/A                                                                      & No                                                              \\ 
                          & Case 94                                                 & 15                                                      & subprocess.Popen()                                                                & Yes                                                     & 18.71                                                               & No                                                                           & \begin{tabular}[c]{@{}c@{}}TypeError: \\ WorkerTimerArgs() \\ takes no arguments\end{tabular} & Yes                                                                       & Yes                                                         \\ 
                          & Case 95                                                 & 10                                                      & subprocess.Popen()                                                                & Yes                                                     & 17.53                                                               & Yes                                                                          & N/A                                                                                           & N/A                                                                      & No                                                              \\ 
                          & Case 96                                                 & 9                                                       & subprocess.Popen()                                                                & Yes                                                     & 15.18                                                               & Yes                                                                          & N/A                                                                                           & N/A                                                                      & Yes                                                             \\
                          & Case 97                                                 & 10                                                      & subprocess.Popen()                                                                & Yes                                                     & 13.81                                                               & Yes                                                                          & N/A                                                                                           & N/A                                                                      & Yes                                                             \\ 
                          & Case 98                                                 & 11                                                      & subprocess.Popen()                                                                & Yes                                                     & 13.59                                                               & Yes                                                                          & N/A                                                                                           & N/A                                                                      & Yes                                                             \\ 
                          & Case 99                                                 & 58                                                      & subprocess.Popen()                                                                & No                                                      & 9.92                                                                & N/A                                                                          & N/A                                                                                           & N/A                                                                      & No                                                              \\ 
                          & Case 100                                                & 20                                                      & subprocess.Popen()                                                                & Yes                                                     & 21.57                                                               & Yes                                                                          & N/A                                                                                           & N/A                                                                      & No                                                              \\
                          & Case 101                                                & 11                                                      & subprocess.Popen()                                                                & Yes                                                     & 18.75                                                               & Yes                                                                          & N/A                                                                                           & N/A                                                                      & Yes                                                             \\ \hline
\end{tabular}
}
\end{table}